\documentclass[preprint,showpacs,preprintnumbers,amsmath,amssymb]{revtex4}
                                                                                
\usepackage{graphicx}
\usepackage{dcolumn}
\usepackage{bm}

\begin{document}

\title{ Supersymmetric pairing of kinks  for polynomial  
nonlinearities} 

\author{H. C. Rosu\footnote{hcr@ipicyt.edu.mx \hfill {\tiny facdec.tex}} 
and   O. Cornejo-P\'erez}

\affiliation{Potosinian Institute of Science and Technology,\\
Apdo Postal 3-74 Tangamanga, 78231 San Luis Potos\'{\i}, Mexico} 
                                                                                
\date{\today}  

\centerline{ArXiv: math-ph/0401040}

\begin{abstract}
\noindent
We show how one can obtain kink solutions of ordinary differential equations 
with polynomial nonlinearities by an efficient factorization procedure directly
related to the factorization of their nonlinear polynomial part. We focus on reaction-diffusion equations in the travelling frame
and damped-anharmonic-oscillator equations.
We also report an interesting pairing of the kink solutions, a result obtained by reversing the factorization
brackets in the supersymmetric quantum mechanical style. In this way, one gets ordinary differential equations with a
different polynomial nonlinearity possessing kink solutions 
of different width but propagating at the same velocity as the kinks
of the original equation. This pairing of kinks could have many applications. We illustrate
the mathematical procedure with several important cases, among which
the generalized Fisher equation, the FitzHugh-Nagumo equation, and the polymerization fronts
of microtubules.
\end{abstract}  



                                                                                
\maketitle

\section{Introduction}

Factorization of second-order linear differential equations,
such as the Schr\"odinger equation, is a well established method to get
solutions in an algebraic manner \cite{sih}. Here we are interested in
factorizations of ordinary differential equations (ODE) of the type
\begin{equation}\label{e0} 
u^{\prime\prime} +\gamma u^{\prime} + F(u)
= 0~, 
\end{equation} 
where $F(u)$ is a given polynomial in $u$. If the
independent variable is the time then $\gamma$ is a damping constant
and we are in the case of nonlinear damped oscillator equations. Many
examples of this type are collected in the Appendix of a paper of
Tuszy\'nski {\em et al} \cite{tod}. However, the coefficient $\gamma$
can also play the role of the constant velocity of a traveling front if the
independent variable is a traveling coordinate used to reduce a
reaction-diffusion (RD) equation to the ordinary differential form as briefly sketched in the 
following. These RD travelling fronts or kinks are important objects in low dimensional nonlinear phenomenology
describing topologically-switched configurations in many areas of biology, ecology, chemistry and physics. 


Consider a scalar RD equation for $u(x,t)$ 
\begin{equation}\label{e1}
\frac{\partial u}{\partial t}={\cal D}\frac{\partial ^2u}{\partial
x^2}+sF(u)~, 
\end{equation} 
where ${\cal D}$ is the diffusion constant
and $s$ is the strength of the reaction process. Eq~(\ref{e1}) can be
rewritten as 
\begin{equation}
\label{e2}
\frac{\partial u}{\partial t}=\frac{\partial ^2u}{\partial x^2}+F(u)~,
\end{equation} 
where the coefficients have been
eliminated by the rescalings $\tilde{t}=st$ and $\tilde{x}=(s/{\cal D})^{1/2}x$, and
dropping the tilde.  Usually, the scalar RD equation possesses
travelling wave solutions $u(\xi)$ with $\xi =x-{\rm v} t$, propagating at speed
${\rm v}$. For this type of solutions the RD equation turns into the ODE 
\begin{equation}\label{4} 
u^{\prime\prime} +{\rm v} u^{\prime} + F(u) = 0~, 
\end{equation} 
where $'=D=\frac{d}{d\xi}$. 
The latter equation has the same form as nonlinear damped oscillator
equations with the velocity playing the role of the friction constant.

For applications in physical optics and acoustics it is convenient to write the travelling coordinate in the form $\xi = kx-\omega t=k(x-{\rm v}t)$
with $k{\rm v} = \omega$. This is a simple scaling by $k$ of the previous coordinate turning Eq.~(\ref{4}) into the form
\begin{equation}\label{4bis} 
u^{\prime\prime} +\frac{{\rm v}}{k} u^{\prime} + \frac{1}{k^2}F(u) = 0 
\end{equation} 
that can be changed back to the form of Eq.~(\ref{e0}) by redefining $\tilde{\gamma}=\frac{{\rm v}}{k}$ and $\tilde{F}(u)=\frac{1}{k^2}F(u)$.

In general, performing the factorization of Eq.~(\ref{e0}) means the following 
\begin{equation}\label{6}
\Big[D-f_{2}(u)\Big]\Big[D-f_{1}(u)\Big]u=0~.  
\end{equation}
This leads to the equation
\begin{equation}\label{7}
u^{\prime\prime}-\frac{df_{1}}{du}uu^{\prime}-f_{1}u^{\prime}-f_{2}u^{\prime}+f_{1}f_{2}u=0~.
\end{equation}

The following groupings of terms are possible related to different factorizations:
                               
{\em  a) Berkovich grouping}: In 1992, Berkovich \cite{ber} proposed to group the terms as follows
\begin{equation}\label{b1}
u^{\prime\prime}-\left(f_{1}+f_{2}\right)u^{\prime}
+\left(f_{1}f_{2}-\frac{df_{1}}{du}u^{\prime}\right)u=0~, 
\end{equation}
and furthermore discussed a theorem according to which any factorization of an ODE of the form given in Eq.~(\ref{6}) allows to find a class of solutions that can be obtained from 
solving the first-order differential equation 
$u^{\prime}=f_{1}u$. Substituting the latter expression in the Berkovich grouping one gets
\begin{equation}\label{b2}
u^{\prime\prime}-\left(f_{1b}+f_{2b}\right)u^{\prime}
+\left(f_{1b}f_{2b}-\frac{df_{1b}}{du}f_{1b}u\right)u=0
\end{equation}
where we redefined $f_1=f_{1b}$ and $f_2=f_{2b}$ to distinguish this case from our proposal following next.
For the specific form of the ODEs we consider here, Berkovich's conditions read
\begin{equation}\label{bco2}
f_{1b}\left(-\gamma -f_{1b}-\frac{df_{1b}}{du}u\right)=\frac{F(u)}{u}~,
\end{equation}
\begin{equation}\label{bco1}
f_{1b}+f_{2b}=-\gamma ~.
\end{equation}

{\em b) Grouping of this work}: We propose here the different grouping of terms
\begin{equation}\label{og1}
u^{\prime\prime}-\left(\frac{d\phi _{1}}{du}u
+\phi _{1}+\phi _{2}\right)u^{\prime}+\phi _{1}\phi _{2}u=0
\end{equation}
that can be considered the result of changing the Berkovich factorization by setting $f_{1b}=\phi _1$ and $f_{2b}\rightarrow \phi _{2}$ under the conditions
\begin{equation}\label{co1}
\phi _{1}\phi _{2}=\frac{F(u)}{u}~, 
\end{equation}
\begin{equation}\label{co2}
\phi _{1}+\phi _2+\frac{d\phi _{1}}{du}u=-\gamma~.
\end{equation}

The following simple relationship exists between the
factoring functions: 
$$
\phi_{2}=f_{2b}-\frac{df_{1b}}{du}u$$ 
and further (third, and so forth) factorizations can be obtained through linear combinations
of the functions $f_{1b}$, $f_{2b}$ and $\phi _{2}$.

Based on our experience, we think that the grouping we propose is more
convenient than that of Berkovich and also of other people employing
more difficult procedures.  
The main advantage resides in the fact that whereas in Berkovich's scheme Eq.~(\ref{bco2}) is still a differential equation to be solved,
in our scheme we make a choice of the factorization functions by merely factoring polynomial expressions according to Eq.~(\ref{co1}) and 
then imposing Eq.~(\ref{co2}) leads easily to an $n$-depending $\gamma$ coefficient for which the 
factorization works. This fact makes our approach extremely efficient in finding particular solutions of the kink type as one can see in the following.

We will show next on the explicit case of the generalized Fisher
equation all the mathematical constructions related to the
factorization brackets and their supersymmetric quantum mechanical like reverse factorization. In addition, in less 
detail, we treat within the same approach, damped nonlinear 
oscillators of Dixon-Tuszy\'nski-Otwinowski type and the 
FitzHugh-Nagumo equation.

\section{Generalized Fisher equation} 

Let us consider the generalized Fisher equation given by
\begin{equation} 
\label{c1-1} 
u^{\prime\prime}
+\gamma u^{\prime} + u(1-u^n) = 0,
\end{equation}
The case $n=1$ refers to the common Fisher
equation and it will be shortly discussed as a subcase.
Eq. (\ref{co1}) allows to factorize the polynomial function
\begin{equation} \label{c1-2} 
\phi _{1}\phi
_{2}=\frac{F(u)}{u}=(1-u^n)=(1-u^{n/2})(1+u^{n/2}),~ 
\end{equation} 
Now, by choosing 
\begin{equation}\label{c1-3}
\phi _{1}=a_{1}(1-u^{n/2}),\;\; \phi _{2}=\frac{1}{a_{1}}(1+u^{n/2})~,\quad a_1\neq 0~,
\end{equation} 
the explicit forms of $a_{1}$ and
$\gamma$ can be obtained from Eq. (\ref{co2})
\begin{equation} \label{c1-4} 
\frac{d\phi _{1}}{du}u+\phi _{1}+\phi
_{2}=-\frac{n}{2}a_{1}u^{n/2}+a_{1}(1-u^{n/2})+(1/a_{1})(1+u^{n/2})=-\gamma
\end{equation} 
Introducing the notation $h_n=(\frac{n}{2}+1)^{1/2}$ one gets
\begin{equation} \label{c1-5} 
a_{1}=\pm h_{n}^{-1}~,  \qquad 
\gamma=\mp\left( h_{n} + h_{n}^{-1}\right)~.  
\end{equation} 
Then Eq. (\ref{c1-1}) becomes 
\begin{equation} \label{c1-6} 
u^{\prime\prime}
\pm \left(h_{n} + h_{n}^{-1}\right)   u^{\prime} + u(1-u^n) = 0 
\end{equation}
and the corresponding factorization is
\begin{equation} \label{c1-6b} 
\Big[ D \pm
h_n(u^{n/2}+1) \Big] \Big[ D \mp
h_{n}^{-1}(u^{n/2}-1) \Big]u=0~. 
\end{equation} 
It follows that Eq.~(\ref{c1-6}) is
compatible with the first-order differential equation
\begin{equation}\label{c1-7} 
u^{\prime}\mp
h_{n}^{-1}\left(u^{n/2}-1\right)u=0~.  
\end{equation}
Integration of Eq. (\ref{c1-7}) gives for $\gamma >0$
\begin{equation} \label{c1-8} 
u_{>}^{\pm}= \left(1\pm
\exp\Big[\left( h_{n} - h_{n}^{-1}\right) (\xi-\xi _{0})\Big]\right)^{-2/n}~.  
\end{equation}
Rewritten in the hyperbolic form, we get 
\begin{eqnarray} \label{c1-9}
u_{>}^{+}=\left(\frac{1}{2}-\frac{1}{2}\tanh\Big[\frac{1}{2} \left( h_{n} - h_{n}^{-1}\right) (\xi-\xi _{0})\Big]\right)^{2/n}~, \nonumber\\
u_{>}^{-} = \left(\frac{1}{2}-\frac{1}{2}\coth\Big[\frac{1}{2} \left( h_{n} - h_{n}^{-1}\right) (\xi-\xi _{0})\Big]\right)^{2/n}~.
\end{eqnarray} 
The tanh form is precisely the solution
obtained long ago by Wang \cite{wang88} and Hereman and Takaoka
\cite{ht90} by more complicated means. 

Moreover, a different solution is possible for $\gamma <0$
\begin{equation}\label{gfi10}
u_{<}^{\pm}= \left(1\pm
\exp\Big[-  \left( h_{n} - h_{n}^{-1}\right) (\xi-\xi _{0})\Big]\right)^{-2/n},
\end{equation}
or
\begin{eqnarray}
u_{<}^{+}=\left(\frac{1}{2}+\frac{1}{2}\tanh\Big[-\frac{1}{2} \left( h_{n} - h_{n}^{-1}\right) (\xi-\xi _{0})\Big]\right)^{2/n},\nonumber\\
u_{<}^{-} = \left(\frac{1}{2}+\frac{1}{2}\coth\Big[-\frac{1}{2} \left( h_{n} - h_{n}^{-1}\right) (\xi-\xi _{0})\Big]\right)^{2/n}~,
\label{gfi11}
\end{eqnarray} 
respectively.

{\em  2.1 Reversion of factorization brackets without the change of the scaling factors}

Choosing now $\phi _{1}=a_{1}(1+u^{n/2})$ and $\phi_{2}=\frac{1}{a_{1}}\left(u^{n/2}-1\right)$
leads to the same equation (\ref{c1-6}) 
but now with the factorization 
\begin{equation}
\Big[ D \mp h_{n}(u^{n/2}-1) \Big] \Big[ D \pm
h_{n}^{-1}(u^{n/2}+1) \Big]u=0 \label{gfi14}~,
\end{equation}
and therefore the compatibility is with the different first-order equation
\begin{equation}\label{gfi15}
u^{\prime}\pm h_{n}^{-1}\left(u^{n/2}+1\right)u=0~.
\end{equation}
However, the direct integration gives the solution (for $\gamma >0$)
\begin{equation}\label{gfi16}
u=\left(-\frac{1}{1\pm
\exp[\left( h_{n} - h_{n}^{-1}\right) (\xi-\xi _{0})]}\right)^{2/n}=(-1)^{2/n}
\left(1\pm
\exp\Big[\left( h_{n} - h_{n}^{-1}\right) (\xi-\xi _{0})\Big]\right)^{-2/n}~,
\end{equation}
which are similar to the known solution Eq.~(\ref{c1-8}). For  $\gamma <0$, solutions of the type given by Eq.~(\ref{gfi10}) are obtained.

{\em 2.2 Direct reversion of factorization brackets}

Let us perform now a direct inversion of the factorization brackets in (\ref{c1-6b}) similar to what is done in supersymmetric quantum mechanics
in order to enlarge the class of exactly solvable quantum hamiltonians
\begin{equation}\label{gfi22}
\Big[ D \mp h_{n}^{-1}(u^{n/2}-1) \Big]\Big[ D \pm
h_n(u^{n/2}+1) \Big] u=0~. 
\end{equation}
Doing the product of differential operators the following
RD equation is obtained
\begin{equation}\label{gfi23}
u^{\prime\prime} \pm \left( h_{n} + h_{n}^{-1}\right)  u^{\prime} +
u\left[1+u^{n/2}\right]\left[1-h_{n}^{4}u^{n/2}\right] = 0~. 
\end{equation}
Eq. (\ref{gfi23}) is compatible with the equation
\begin{equation}\label{gfi24}
u^{\prime}\pm h_n\left(u^{n/2}+1\right)u=0,
\end{equation}
and integration of the latter gives the kink solution of Eq.~(\ref{gfi23})
\begin{equation}\label{gfi26}
u_{>}^{\pm}=\left(-\frac{1}{1\pm
\exp[(h_n^3-h_n)(\xi-\xi _{0})]}\right)^{\frac{2}{n}}=
\left(1\pm
\exp\Big[(h_n^3-h_n)(\xi-\xi _{0})\Big]\right)^{-\frac{2}{n}}
\end{equation}
for $\gamma >0$. On the other hand,  for $\gamma <0$ the exponent is the same but of opposite sign.
Hyperbolic forms of the latter solutions are easy to write down and are similar up to widths to Eqs.~(\ref{c1-9}) and (\ref{gfi11}), respectively.

Thus, a different RD equation given by (\ref{gfi23}) 
with modified polynomial terms and its solution have been found by reverting
the factorization terms of Eq. (\ref{c1-6}). Although the reaction polynomial is different the velocity parameter remains the same.
This is the main result of this work: {\em At the velocity corresponding to the travelling kink of a given RD equation there is another 
propagating kink corresponding to a different RD equation that is related to the original one by reverse factorization}. 
We can call this kink as the supersymmetric (susy) kink because of the mathematical construction.

Finally, one can ask if the process of reverse factorization can be continued with Eq.~(\ref{gfi23}). It can be shown that this is not the case because Eq.~(\ref{gfi23})
has already a discretized (polynomial-order-dependent) $\gamma$ and this fact prevents further solutions of this type. Suppose we consider the following factorization functions 
\begin{equation}\label{f1}
\tilde{\phi}_1=\tilde{a}_{1}^{-1}\Big[1-h_n^4u^{n/2}\Big]~, \qquad \tilde{\phi} _2=\tilde{a}_1\left(1+u^{n/2}\right)~.
\end{equation}
Then, one gets $\tilde{a}_1=\pm h_{n}^{3}$ and solve $\tilde{a}_1^{-1}+\tilde{a}_1=h_{n}^{-1}+h_n$.
The solutions are: $n=0$, which implies linearity, and $n=-4$, which leads to a Milne-Pinney equation.
On the other hand, Eq.~(\ref{gfi23}) with an arbitrary $\gamma$ can be treated by the 
inverse factorization procedure to get the susy partner RD equation and its susy kink.

{\em 2.3 Subcase $n=1$} 

This subcase is the original Fisher equation describing the propagation of mutant genes
\begin{equation}\label{Fish-1}
\frac{\partial u}{\partial t}=\frac{\partial^2 u}{\partial
x^2}+u(1-u)~.
\end{equation}

In the travelling frame, the Fisher equation has the form
\begin{equation} 
\label{Fish0} 
u^{\prime\prime}
+\alpha u^{\prime} + u(1-u) = 0~.
\end{equation}
When the $\gamma$ parameter takes the value $\gamma _1=\frac{5}{6} \sqrt{6}$ (i.e., $h_1=\frac{\sqrt{6}}{2}$) one can factor Fisher's equation 
and employing our method leads easily to the known kink solution  
\begin{equation}\label{Fish1}
u_{\rm F}=\frac{1}{4}\left(1-\tanh\Big[\frac{\sqrt{6}}{12}(\xi-\xi _{0})\Big]\right)^{2}~
\end{equation}
that was first obtained by Ablowitz and Zeppetella \cite{az79} with a series solution method.
On the other hand, the susy kink for this case reads
\begin{equation}\label{Fish2}
u_{\rm F, susy}=\frac{1}{4}\left(1-\tanh\Big[\frac{\sqrt{6}}{8}(\xi-\xi _{0})\Big]\right)^{2}~,
\end{equation}
i.e., it has a width one and a half times greater than the common Fisher kink and is a solution of the partner equation
\begin{equation} \label{Fish3}
u^{\prime\prime}+\frac{5\sqrt{6}}{6} u^{\prime} + u\left(1-\frac{5}{4}u^{1/2}-\frac{9}{4}u\right) = 0~.
\end{equation}

A plot of the kinks $u_{\rm F}$ and $u_{\rm F,susy}$ is displayed in Fig.~1.
%
\begin{figure}[x]
\centerline{\includegraphics[scale=0.6]{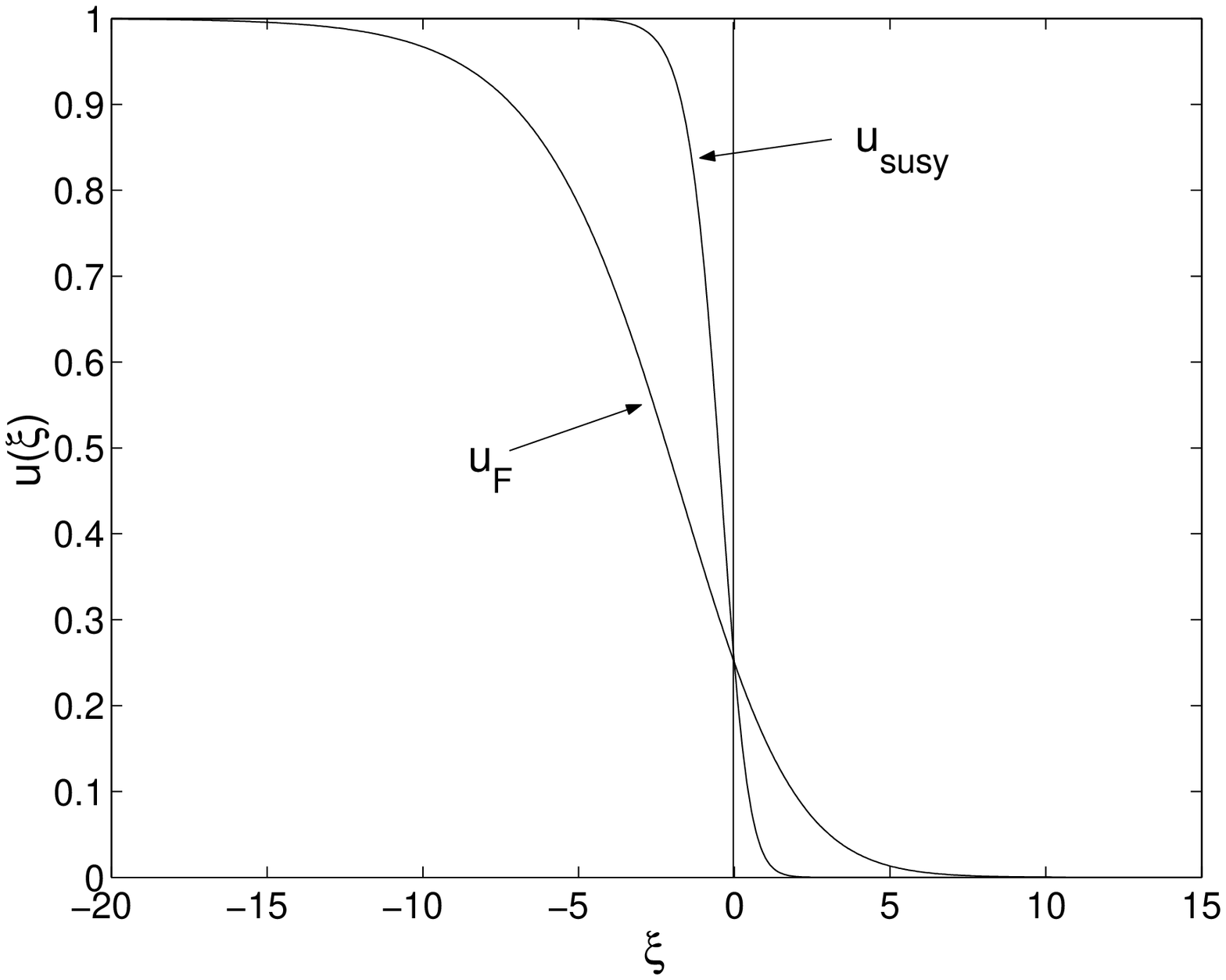}}    
\caption{ The front of mutant genes (Fisher's wave of advance) in
a population and the partner susy kink propagating with the same
velocity. The axis are in arbitrary units.}  
\end{figure} 

{\em  2.4 Subcase $n=6$}

This subcase is of interest in the light of experiments on polymerization patterns of microtubules in centrifuges.
It has been discovered that the polymerization of the tubulin dimers  
proceeds in a kink-switching fashion propagating with a constant velocity within the sample. 
Portet, Tuszynski and Dixon \cite{p} used RD equations to discuss the modification of self-organization patterns of MTs as well as the tubulin polymerization under the influence of reduced gravitational fields. They used the value $n=6$ for the mean critical number of tubulin dimers at which the polymerization process starts and showed that the same nucleation number enters the polynomial term of the RD process
for the number concentration ${\rm c}$ of tubulin dimers 
\begin{equation}\label{nconc}
{\rm c}^{\prime\prime} +\frac{5}{2}{\rm c}^{\prime} +
{\rm c}\left(1-{\rm c}^{6}\right)= 0~. 
\end{equation}
The polymerization kink in their work reads
\begin{equation}\label{polykink}
{\rm c}_{{\rm PTD}}     
=2^{-\frac{1}{3}}\left(1-\tanh\Big[\frac{3}{4}(\xi-\xi _{0})\Big]\right)^{1/3}~.
\end{equation}
On the other hand, the susy polymerization kink (see Fig.~(2)) of the form
\begin{equation}\label{polyk1}
{\rm c}_{\rm susy}   
=
2^{-\frac{1}{3}}\Big(1-\tanh[3(\xi-\xi _{0})]\Big)^{1/3}
\end{equation}
can be taken into account according to the hyperbolic form of Eq.~(\ref{gfi26}). It propagates with the same speed and corresponds to the equation 
\begin{equation}\label{nconc1}
{\rm c}^{\prime\prime} \pm\frac{5}{2}{\rm c}^{\prime} +
{\rm c}\left(1-15 {\rm c}^{3}-16{\rm c}^{6}\right)= 0~. 
\end{equation}
In principle, this equation could be obtained as a consequence of modifying the kinetics steps in 
the microtubule polymerization process.  

\begin{figure}[x]
\centerline{\includegraphics[scale=0.6] {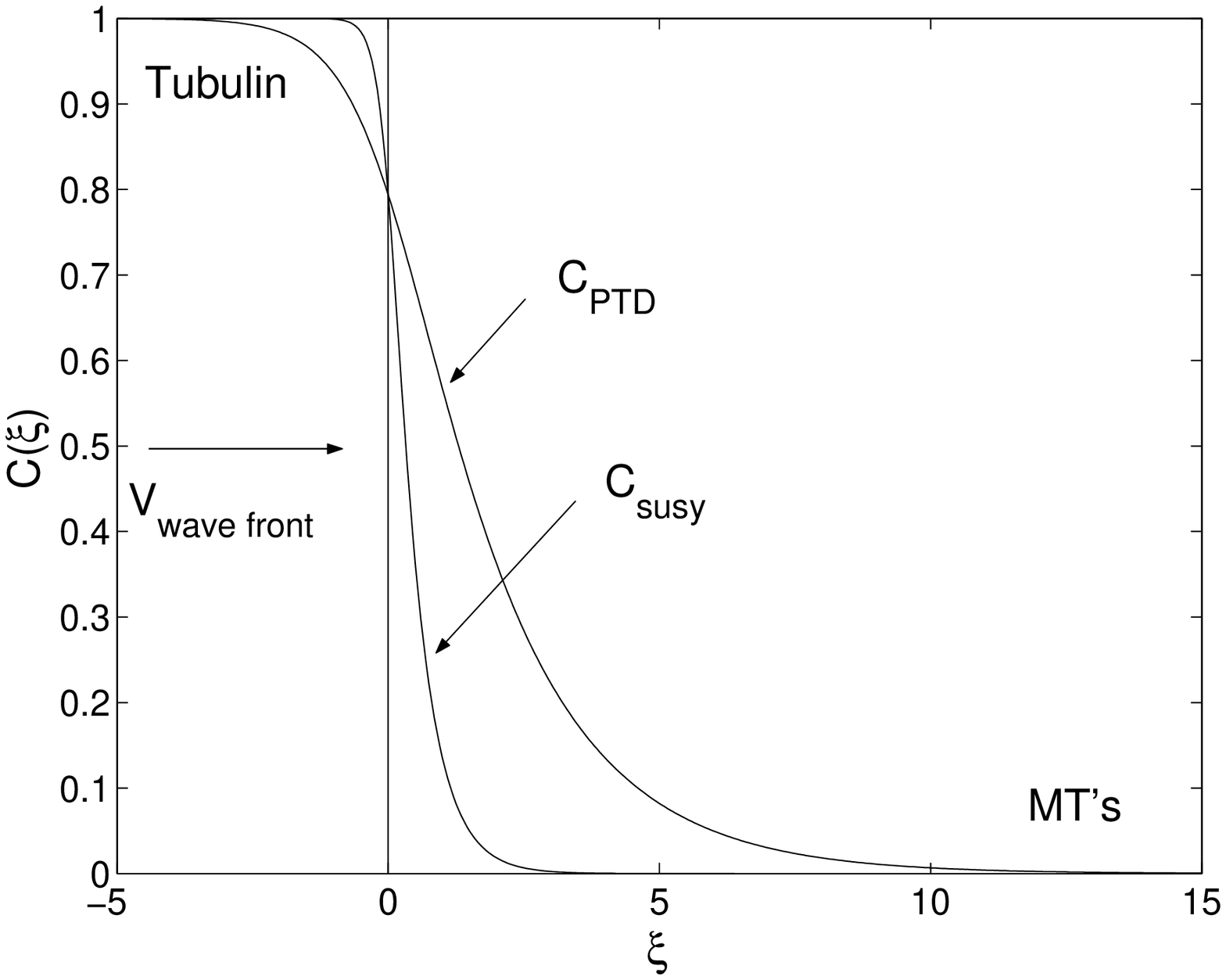}}    
\caption{ The polymerization kink of Portet, Tuszy\'nski and Dixon \cite{p} and  the susy kink propagating with the same velocity (axis in arbitrary units).} 
\end{figure} 


\section{Equations of the Dixon-Tuszy\'nski-Otwinowski type}

In the context of damped anharmonic oscillators, Dixon {\em et al} \cite{dto91} studied equations of the type (in this section, we use $'=D_{\tau}=\frac{d}{d\tau}$)
\begin{equation}\label{dto1}
u^{\prime\prime}+u^{\prime}+Au-u^{n-1}\equiv u^{\prime\prime}+u^{\prime}+u(\sqrt{A}-u^{\frac{n}{2}-1})(\sqrt{A}+u^{\frac{n}{2}-1}) =0
\end{equation}
and gave solutions for the cases $A=\frac{2}{9}$ and $A=\frac{3}{16}$, with $n=4$ and $n=6$, respectively. 
For this case, time is the independent variable. The factorization method works nicely if one uses $g_n=\sqrt{n/2}$ and dealing
with the more general equation
\begin{equation}
u^{\prime\prime}\pm\sqrt{A}(g_n+g_{n}^{-1}) 
u^{\prime}+u(A-u^{n-2})=0, \label{otwi1}
\end{equation}
for which we can employ either the factorization functions
\begin{eqnarray}
\phi _{1}=\mp g_{n}^{-1}\left(\sqrt{A}-u^{\frac{n}{2}-1}\right),\quad
\phi _{2}=\mp g_n\left(\sqrt{A}+u^{\frac{n}{2}-1}\right)\label{otwi2}
\end{eqnarray}
or
\begin{eqnarray}
\phi _{1}=\mp g_{n}^{-1}\left(\sqrt{A}+u^{\frac{n}{2}-1}\right),\quad
\phi _{2}=\mp g_n\left(\sqrt{A}-u^{\frac{n}{2}-1}\right)~.\label{otwi3}
\end{eqnarray}
Then, Eq. (\ref{otwi1}) can be factored in the forms
\begin{equation}
\Big[ D_{\tau} \pm g_n(u^{\frac{n}{2}-1}+\sqrt{A})
\Big]\Big[ D_{\tau} \mp g_{n}^{-1}(u^{\frac{n}{2}-1}-\sqrt{A})
\Big] u=0 \label{otwi4}
\end{equation}
and
\begin{equation}
\Big[D_{\tau} \mp g_n(u^{\frac{n}{2}-1}-\sqrt{A})
\Big]\Big[D_{\tau} \pm g_{n}^{-1}(u^{\frac{n}{2}-1}+\sqrt{A})
\Big] u=0~. \label{otwi5}
\end{equation}
Thus, Eq. (\ref{otwi1}) is compatible with the equations
\begin{eqnarray}
 u^{\prime} \mp g_{n}^{-1}\left(u^{\frac{n}{2}-1}-\sqrt{A}\right)
 u=0,\label{otwi6}\\
u^{\prime} \pm g_{n}^{-1}\left(u^{\frac{n}{2}-1}+\sqrt{A}\right) u=0
\label{otwi7}
\end{eqnarray}
that follows from Eq.~(\ref{otwi4}) and Eq.~(\ref{otwi5}).
Integration of Eqs. (\ref{otwi6}), (\ref{otwi7}) gives the
solution of Eq. (\ref{otwi1})
\begin{equation}
u_{>}=\left(\frac{\sqrt{A}}{1\pm
\exp\Big[\sqrt{A}(g_n -g_{n}^{-1})(\tau-\tau_{0})\Big]}\right)^{\frac{2}{n-2}},\quad\quad
\gamma >0 \label{otwi8}
\end{equation}
and
\begin{equation}
u_{<}=\left(\frac{\sqrt{A}}{1\pm
\exp\Big[-\sqrt{A}(g_n - g_{n}^{-1})(\tau-\tau_{0})\Big]}\right)^{\frac{2}{n-2}},
\quad\quad\gamma <0~.\label{otwi9}
\end{equation}
The solutions obtained by Dixon {\em et al} are particular cases of the latter formulas.

Reversing now the factorization brackets in (\ref{otwi4})
\begin{equation}
\Big[ D_{\tau} \mp g_{n}^{-1}\left(u^{\frac{n}{2}-1}-\sqrt{A}\right)
\Big]\Big[ D_{\tau} \pm g_n(u^{\frac{n}{2}-1}+\sqrt{A})
\Big] u=0 \label{otwi10}
\end{equation}
leads to the following equation
\begin{equation}
u^{\prime\prime}\pm\sqrt{A}(g_n +g_{n}^{-1})
u^{\prime}+u\left(\sqrt{A}+u^{\frac{n}{2}-1}\right)\left(\sqrt{A}-\frac{n^2}{4}u^{\frac{n}{2}-1}\right)=0,
\label{otwi11}
\end{equation}
which is compatible with the equation
\begin{equation}
u^{\prime} \pm g_n\left(u^{\frac{n}{2}-1}+\sqrt{A}\right)u=0
\label{otwi12}
\end{equation}
whose integration gives the solution of Eq.~(\ref{otwi11})
\begin{equation}
u_{>}= \left(\frac{\sqrt{A}}{1\pm
\exp[\sqrt{A}g_n(\tau-\tau_{0})]}\right)^{\frac{2}{n-2}},\quad\quad\gamma >0\label{otwi13}
\end{equation}
and
\begin{equation}
u_{<}= \left(\frac{\sqrt{A}}{1\pm
\exp[-\sqrt{A}g_n(\tau-\tau_{0})]}\right)^{\frac{2}{n-2}},\quad\quad\gamma <0~.\label{otwi14}
\end{equation}

\section{FitzHugh-Nagumo Equation}

Let us consider the FitzHugh-Nagumo equation
\begin{equation}\label{fhn0}
\frac{\partial u}{\partial t}-\frac{\partial^2 u}{\partial
x^2}+u(1-u)(a-u)=0~,
\end{equation}
where $a$ is a real constant. If $a=-1$, one gets the real Newell-Whitehead equation describing the dynamical behaviour near the bifurcation point for
the Rayleigh-B\'enard convection of binary fluid mixtures.
The travelling frame form of (\ref{fhn0}) has been discussed in detail by Hereman and Takaoka \cite{ht90} 
\begin{equation}
u^{\prime\prime}+\gamma u^{\prime}+u(u-1)(a-u)=0. \label{fhn1}
\end{equation}
The FitzHugh-Nagumo polynomial function allows the following factorizations:
\begin{eqnarray}
\phi _{1}=\pm(\sqrt{2})^{-1}(u-1),\quad
\phi _2=\pm\sqrt{2}(a-u)\label{fhn2}
\end{eqnarray}
when the $\gamma$ parameter is equal to $\gamma _{a1}=\pm\frac{-2a+1}{\sqrt{2}}$ that we also write as $\gamma _{a,1}=\pm \sqrt{a}(g_{a1}-g_{a1}^{-1})$, where $g_{a1}=-\sqrt{2a}$.

In addition, we can employ the factorization functions
\begin{eqnarray}
\phi _{1}=\pm(\sqrt{2})^{-1}(a-u),\quad
\phi _2=\pm\sqrt{2}(u-1)\label{fhn3}
\end{eqnarray}
when $\gamma _{a,2}=\pm\frac{-a+2}{\sqrt{2}}$, or written again in the more symmetric form $\gamma _{a,2}=\pm \sqrt{a}(g_{a2}-g_{a2}^{-1})$, where $g_{a2}=-\sqrt{a/2}$.  Thus, Eq. (\ref{fhn1}) can
be factored in the two cases
\begin{equation}
u^{\prime\prime}\pm \gamma _{a,1} 
u^{\prime}+u(u-1)(a-u)=0, \label{fhn3a}
\end{equation}
and
\begin{equation}
u^{\prime\prime}\pm \gamma _{a,2} 
u^{\prime}+u(u-1)(a-u)=0~.
\label{fhn3b}
\end{equation}
In passing, we notice that for the Newell-Whitehead case $a=-1$ the two equations coincide and are the same as the generalized Fisher equation for $n=2$.

In factorization bracket forms, Eqs. (\ref{fhn3a}) and (\ref{fhn3b}) are written as follows
\begin{equation}
\Big[ D \mp\sqrt{2}(a-u)\Big]\Big[ D \pm
(\sqrt{2})^{-1}(1-u)\Big] u=0 \label{fhn4}
\end{equation}
and
\begin{equation}
\Big[ D \mp \sqrt{2}(u-1)\Big]\Big[ D \mp
(\sqrt{2})^{-1}(a-u)\Big] u=0, \label{fhn5}
\end{equation}
and are compatible with the first order differential equations
\begin{eqnarray}
u^{\prime} \pm (\sqrt{2})^{-1}(1-u)u=0,\quad\quad {\rm for} \,\gamma  _{a,1}~,
 \label{fhn6} \\
u^{\prime} \mp (\sqrt{2})^{-1}(a-u)u=0,\quad\quad {\rm for} \, \gamma _{a,2}~.
 \label{fhn7}
\end{eqnarray}
Integration of the latter equations 
gives the
solution of Eq. (\ref{fhn1}) for the two different values of the
wave front velocity $\gamma _{a,1}$ and $\gamma _{a,2}$.

For Eq.~(\ref{fhn3a}) we get 
\begin{eqnarray}
u_{>}= \frac{1}{1\pm
\exp[(\sqrt{2})^{-1}(\xi-\xi _{0})]},
\quad u_{<}=
\frac{1}{1\pm
\exp[-(\sqrt{2})^{-1}(\xi-\xi _{0})]},
\end{eqnarray}
for $\gamma _{a,1}$ positive and negative, respectively.

As for Eq.~(\ref{fhn3b}), the solutions are
\begin{eqnarray}
u_{>}= \frac{a}{1\pm
\exp[-(\sqrt{2})^{-1}a(\xi -\xi _{0})]},
\quad u_{<}=
\frac{a}{1\pm
\exp[(\sqrt{2})^{-1}a(\xi -\xi _{0})]},
\end{eqnarray}
for $\gamma _{a,2}$ positive and negative, respectively.

Considering now the factorizations (\ref{fhn4}) and (\ref{fhn5}), the
change of order of the factorization brackets gives
\begin{equation}
\Big[ D \pm (\sqrt{2})^{-1}(1-u)\Big] \Big[ D
\mp\sqrt{2}(a-u)\Big]u=0 \label{fhn8}
\end{equation}
and
\begin{equation}
\Big[ D \mp (\sqrt{2})^{-1}(a-u)\Big] \Big[ D \mp
\sqrt{2}(u-1)\Big]u=0~. \label{fhn9}
\end{equation}
Doing the product of differential operators (and considering the
factorization term $u^{\prime}- \phi _{2}u=0$) gives the following RD
equations
\begin{equation}
u^{\prime\prime} \pm \gamma _{a1}    
u^{\prime} + u(4u-1)(a-u)= 0, \label{fhn10}
\end{equation}
and
\begin{equation}
u^{\prime\prime} \pm \gamma _{a2}    
u^{\prime} +u(u-1)(a-u-3u^2)= 0, \label{fhn11}
\end{equation}
Eqs. (\ref{fhn10}) and (\ref{fhn11}) are compatible with the
equations
\begin{equation}
u^{\prime} \mp\sqrt{2}(a-u)u=0\label{fhn12}
\end{equation}
and
\begin{equation}
u^{\prime} \mp \sqrt{2}(u-1)u=0~, \label{fhn13}
\end{equation}
respectively. Integrations of Eqs. (\ref{fhn12}) and (\ref{fhn13})
give the solutions of Eqs. (\ref{fhn10}) and (\ref{fhn11}), respectively.
The explicit forms are the following:

(i) for (\ref{fhn10})
\begin{eqnarray}
u_{>}= \frac{a}{1\pm
\exp[-\sqrt{2}a(\xi-\xi _{0})]}~, 
\quad u_{<}=
\frac{a}{1\pm
\exp[\sqrt{2}a(\xi -\xi _{0})]}~. 
\end{eqnarray}

(ii) for (\ref{fhn11})  
\begin{eqnarray}
u_{>}= \frac{1}{1\pm \exp[\sqrt{2}(\xi -\xi _{0})]}~, 
\quad u_{<}= \frac{1}{1\pm
\exp[-\sqrt{2}(\xi -\xi _{0})]}~. 
\end{eqnarray}

\section{Conclusion}

This paper has been concerned with stating an efficient factorization scheme of ordinary differential equations with polynomial nonlinearities that leads to an easy finding 
of analytical solutions of the kink type that previously have been obtained by far more cumbersome procedures. The main result is an interesting pairing between equations with different polynomial
nonlinearities, which is obtained by applying the susy quantum mechanical reverse factorization. The kinks of the two nonlinear equations are of different widths but
they propagate at the same velocity, or if we deal with damped polynomial nonlinear oscillators the two kink solutions correspond to the same friction coefficient.
Several important cases, such as the generalized Fisher and the FitzHugh-Nagumo equations, have been shown to be simple mathematical exercises for this factorization
technique. The physical prediction is that for commonly occurring propagating fronts, there are two kink fronts of different widths at a given propagating velocity.    
Moreover, the reverse factorization procedure can be also applied to the Berkovich scheme with similar results. It will be interesting to apply the approach of this work to the 
discrete case in which various exact results have been obtained in recent years \cite{discrete}. 
More general cases in which the coefficient $\gamma$ is an arbitrary function could also be of much interest because of possible applications.


\end{document}